\documentclass{jpsj2}

\usepackage{amsmath,amsfonts,amsthm,amssymb,bm,cite,eucal}
\usepackage{graphicx}

\def\e{\mathrm{e}}
\def\d{\mathrm{d}}
\def\im{\mathrm{i}}
\def\6{\partial}
\def\f{\Phi}
\def\F{\bm{\Phi}}
\def\<{\langle}
\def\>{\rangle}
\def\N{\mathcal{N}}
\def\T{\mathcal{T}}

\title{{\bf Multicomponent Bright Solitons in $F=2$ Spinor Bose--Einstein Condensates}}

\author{Masaru \textsc{Uchiyama}$^{1}$, 
Jun'ichi \textsc{Ieda}$^{2}$\thanks{ieda@imr.tohoku.ac.jp} 
and Miki \textsc{Wadati}$^{1}$\thanks{wadati@phys.s.u-tokyo.ac.jp}}

\inst{$^{1}$Department of Physics, Graduate School of Science,
University of Tokyo, Tokyo 113-0033 \\
$^{2}$Institute for Materials Research, Tohoku University, Sendai 980-8577}

\abst{We study soliton solutions for the Gross--Pitaevskii equation of 
the spinor Bose--Einstein 
condensates with hyperfine spin $F=2$ in one-dimension. 
Analyses are made in two ways: by assuming single-mode amplitudes and 
by generalizing Hirota's direct method for multi-components. 
We obtain one-solitons of single-peak type 
in the ferromagnetic, polar and cyclic states, respectively. 
Moreover, twin-peak type solitons 
both in the ferromagnetic and the polar state are found. 
}

\kword{$F=2$ Bose--Einstein condensate, hyperfine spin, 
multi-component solitons, 
integrability, direct method, 
multi-component Gross--Pitaevskii equation}

\begin{document}
\maketitle
\newpage

\section{Introduction} 
\label{sec:intro}

Since the realization in experiments, Bose--Einstein condensate (BEC) 
of ultra-cold atoms has attracted much interest of theoretical physicists. 
Under optical dipole traps, the hyperfine spin of atoms
remains to be active and leads to
the spinor BEC \cite{otrap1,otrap2,otrap3,F2MIT,F2Hamb,GTU,F2Gaku,F2CNRS}. 
So far, spinor BECs have been found to show a variety of phases 
\cite{OM,Ho,CYHo,UK,DH}. For the hyperfine spin $F=1$ state, the ground-state phase can be either polar (antiferromagnetic) with $^{23}$Na \cite{otrap1,otrap2} 
or ferromagnetic with $^{87}$Rb \cite{F2Hamb,GTU,F2Gaku,F2CNRS}.
$F=2$ condensates have been realized with both $^{23}$Na \cite{F2MIT}
and $^{87}$Rb \cite{F2Hamb,GTU,F2Gaku,F2CNRS} atomic species.
The $F=2$ BECs are classified into three distinct phases referred
to as ferromagnetic, polar, and ``cyclic" \cite{CYHo,UK}.
However, the ground-state phase of the $F=2$ state at zero magnetic field is under discussion \cite{SU} due to its very short lifetime (a few milliseconds) for which the equilibrium state cannot be reached.
Compared to the $F=1$ cases, the spin dynamics of the $F=2$
BEC is less well-understood, especially in the cyclic phase \cite{PM},
giving rise to experimental and theoretical challenges. 

Recently, solitons of spinor BECs in one-dimension have been studied 
analytically \cite{IedaMW1,IedaMW2,WadatiT,UchiIW-dark,KuroW} 
and numerically \cite{Malomed,Kivshar,You}. 
In experiments, matter-wave dark and bright solitons are
produced only for single-component BEC 
\cite{soliton1,soliton2,soliton3,soliton4}. 
For a generic hyperfine spin $F$, 
the dynamics is described by the $(2F+1)$-component 
Gross--Pitaevskii (GP) equation. 
The one-component GP equation is called 
the nonlinear Schr\"odinger equation (NLSE). 
If all the spin-dependent interactions vanish 
and only the intensity interaction exists, 
the multi-component GP equation is equivalent to 
the vector NLSE, which is also called the Manakov equation. 
The soliton solutions of these systems are well-known 
\cite{ZakSha1,ZakSha2,Manakov,Tsuchida}. 
For $F=1$, at special sets of coupling constants
one-soliton solutions and two-soliton collisions were explicitly shown 
for the bright soliton under the vanishing boundary conditions 
\cite{IedaMW1,IedaMW2}, 
and for the dark soliton \cite{UchiIW-dark} 
and the bright soliton \cite{KuroW}
under the nonvanishing boundary conditions, 
by finding the map from the GP equation to the $2\times 2$ matrix NLSE, 
which can be solved by the inverse scattering method 
\cite{TsuchidaW,IedaUchiW-ISM}. 
However, for higher spins, such a map to a known integrable equation 
has not been found and we need to look for alternative methods. 

In this paper, we aim at seeking higher-spinor BEC solitons in one-dimension. 
For this purpose, we employ two different methods for 
demonstrating $F=2$ BEC bright one-solitons. 
One facile method, the single-mode analysis \cite{WadatiT}, effectively 
utilizes the reduction of the multi-component GP equation to the one-component one. 
Another new method is a generalization of Hirota's direct method 
\cite{Hirota}
to multi-components. 
Hirota's direct method has been successfully applied to get solitons 
in one-component equation systems, 
and we prove its applicability and strength even for the multi-component systems. 
Indeed, this method gives solitons beyond the single-mode analysis. 
Both methods are regarded to be effective even for nonintegrable equations, 
and suitable for the first step of investigation. 
The results include not only ordinary single-peak solitons but also 
twin-peak solitons which cannot be expressed as the superposition of 
two single-peak solitons. 

The paper is organized as follows. 
In Sec. \ref{sec:F=2BEC} the GP equation for the $F=2$ spinor BEC 
is introduced. 
In Sec. \ref{sec:SM} we study the single-mode analysis. 
In Sec. \ref{sec:Hirota} Hirota's direct method is generalized 
for the $F=2$ GP equation. 
In Sec. \ref{sec:soliton} we present one-soliton solutions 
obtained by Hirota's method and discuss their properties. 
The last section is devoted to discussion and conclusion.

\section{$F=2$ Spinor BEC in One-Dimension}
\label{sec:F=2BEC}

In the mean-field theory, the $F=2$ spinor BEC is characterized by 
the local order parameter (or, the macroscopic wavefunction) 
with five components, 
$\bm{\Phi}=(\Phi_2,\Phi_1,\Phi_0,\Phi_{-1},\Phi_{-2})$, 
reflecting the five spin degrees of freedom. 
For the magnetic quantum number $j=-2,\cdots,2$ 
with respect to the quantization axis chosen in the $z$-direction, 
$\Phi_j=\Phi_j(x,t)=\< \hat{\Psi}_j(x,t)\>$. 
In words, $\Phi_j$ 
are given by the ground state expectation value of the boson operators 
$\hat{\Psi}_j(x,t)$, 
which satisfy the equal-time commutation relation 
$[\hat{\Psi}_\alpha(x,t),\hat{\Psi}_\beta^\dagger(x',t)]
=\delta_{\alpha\beta}\delta(x-x')$ 
for $\alpha,\beta=-2,\cdots,2$.

We consider the dynamics of the $F=2$ spinor BEC in one-dimension. 
The evolution equation for the local order parameters is described by 
the multi-component Gross--Pitaevskii (GP) equation, 
\begin{align}
\label{eq:GP}
\im\hbar\frac{\6\F}{\6 t}=\frac{\delta E_{\mathrm{GP}}[\F]}{\delta \F^*}. 
\end{align}
Here the energy functional is defined by \cite{OM,Ho,CYHo,UK}
\begin{align}
\label{eq:GPenergy}
E_{\mathrm{GP}}[\F]=
\int_{-\infty}^\infty \d x
\left(\frac{\hbar^2}{2m}|\6_x\F|^2+\frac{c_0}{2}n^2
+\frac{c_2}{2}\bm{f}^2+\frac{c_4}{2}|\Theta|^2\right).
\end{align}
The coupling constants $c_i$ are real 
and can be expressed in terms of a transverse confinement radius 
and a linear combination of the $s$-wave scattering lengths 
of atoms \cite{IedaMW1}. 
The interaction energy is derived from the short-range interactions 
of atoms in the scattering channel with total spin $0$, $2$, $4$, 
and is given in terms of 
the number density 
\begin{align}
n=\sum_{\alpha=-2,\cdots, 2} \f_\alpha^*\f_\alpha, 
\end{align}
the spin densities $\bm{f}=(f^x,f^y,f^z)$, where for $i=x,y,z$, 
\begin{align}
f^i=\sum_{\alpha,\beta=-2,\cdots, 2} 
\f_\alpha^*\mathsf{f}^i_{\alpha\beta}\f_\beta,
\end{align}
and the singlet-pair amplitude \cite{CYHo,UK}
\begin{align}
\Theta=2\f_2\f_{-2}-2\f_1\f_{-1}+\f_0^2. 
\end{align}
The meaning of $\Theta$ is clear if we write with 
the Clebsch-Gordan coefficient as 
$\Theta=\sqrt{5}\sum_{j,j'}\<00|2j;2j'\> \f_j\f_{j'}$,
\emph{i.e.} it measures the formation of spin-singlet ``pairs" of bosons. 
The prefactor $\sqrt{5}$ is introduced just for convenience. 
The $c_4$-term in (\ref{eq:GPenergy}) includes the scattering process 
$2+(-2)\leftrightarrow 0+0$, which changes the $z$-component of 
the spin states of bosons by two and is absent for $F=1$. 
We also write the spin densities as $f^\pm=f^x\pm\im f^y$. 
The spin matrices $\mathsf{f}^i$ in $F=2$ are explicitly represented as 
\begin{align}
&\mathsf{f}^x=\left[
\begin{array}{ccccc}
0&1&0&0&0\\
1&0&\sqrt{6}/2&0&0\\
0&\sqrt{6}/2&0&\sqrt{6}/2&0\\
0&0&\sqrt{6}/2&0&1\\
0&0&0&1&0
\end{array}
\right],
\qquad
&\mathsf{f}^y=\left[
\begin{array}{ccccc}
0&-\im&0&0&0\\
\im&0&-\im\sqrt{6}/2&0&0\\
0&\im\sqrt{6}/2&0&-\im\sqrt{6}/2&0\\
0&0&\im\sqrt{6}/2&0&-\im\\
0&0&0&\im&0
\end{array}
\right],\nonumber\\
&\mathsf{f}^z=\left[
\begin{array}{ccccc}
2&0&0&0&0\\
0&1&0&0&0\\
0&0&0&0&0\\
0&0&0&-1&0\\
0&0&0&0&-2
\end{array}
\right].
\end{align}
Without the magnetic field, the energy (\ref{eq:GPenergy}) 
is invariant under an SU(2) rotation 
and the system has an SU(2) symmetry. 
In particular, an obvious symmetry is the one under 
$\exp[\im\pi\mathsf{f}^x]:\f_j\mapsto \f_{-j}$. 

We set $\hbar=1$, $2m=1$ to simplify the expressions. 
With an operator $\mathcal{L}=\im\6_t+\6^2_x$, 
the explicit form of GP equation (\ref{eq:GP}) is 
\begin{subequations}
\begin{align}
&\mathcal{L}\f_{\pm 2}=
c_0 n \f_{\pm 2}
+c_2\left(\pm 2f^z\f_{\pm 2} +f^{\mp}\f_{\pm 1}\right)
+c_4\Theta \f_{\mp 2}^*,
\\
&\mathcal{L}\f_{\pm 1}=
c_0 n \f_{\pm 1}
+c_2\left(f^{\pm}\f_{\pm 2} \pm f^z\f_{\pm 1} +\frac{\sqrt{6}}{2}f^{\mp}\f_0\right)
-c_4\Theta \f_{\mp 1}^*,
\\
&\mathcal{L}\f_0=
c_0 n \f_0+c_2\left(
\frac{\sqrt{6}}{2}f^+\f_1+\frac{\sqrt{6}}{2}f^-\f_{-1}
\right)
+c_4\Theta \f_{0}^*.
\end{align}
\label{eq:GPexplicit}
\end{subequations}
The right-hand sides include cubic terms with respect to $\f_j$. 
If the spin-dependent interactions are absent, 
\emph{i.e.} $c_2=c_4=0$, 
the GP equation is reduced to 
the Manakov equation with five components and 
solutions for the initial problem as well as 
multi-solitons are known 
in the formalism of the inverse scattering method \cite{Manakov,Tsuchida}. 
However, this is a trivial reduction for spinor condensates. 
In the presence of the spin-dependent interactions 
where rich phenomena are expected, 
the GP equation becomes highly correlated and hard to be solved explicitly. 

We concentrate on soliton solutions for the $F=2$ GP equation. 
A soliton propagates keeping its own wave properties. 
Through its free translational motion, 
physical quantities given by the integral of densities 
characterize the soliton, 
such as the particle number $N=\int n\d x$, the spin $\bm{F}=\int \bm{f}\d x$ and  
the volume of the singlet-pair $S=\int|\Theta|\d x$. 

In the subsequent sections, we attempt to derive one-soliton solutions 
with non-trivial spin degrees of freedom 
and clarify their physical properties. 
We apply two methods and 
obtain several one-soliton solutions for the GP equation 
(\ref{eq:GPexplicit}).

\section{Single-Mode Analysis}
\label{sec:SM}

The single-mode analysis assumes the following amplitude 
for the order parameters \cite{WadatiT}: 
\begin{align}
\F(x,t)=\bm{A}\phi(x,t),
\end{align}
where $\bm{A}=(A_2,A_1,A_0,A_{-1},A_{-2})$. 
That is, the order parameters have the same spatial profile 
but can have different magnitude. 
Normalization is such that $\sum_j|A_j|^2=1$. 
We require the GP equation to lead to 
the one-component nonlinear Schr\"odinger equation for $\phi$, 
\begin{align}
\im\6_t\phi+\6_x^2\phi-C|\phi|^2\phi=0
\label{eq:1NLSE}
\end{align}
with $C$ being a real constant. 
This imposes the consistency conditions on the nonlinear terms of 
the GP equation. 
By the freedom of SU(2) rotation, we can fix the spin in the $z$-direction 
to have $f^+=f^-=0$. 
Then the conditions read 
\begin{align}
E_{j_1}=\cdots =E_{j_k},
\end{align}
for $j_1,\cdots,j_k\in\{-2,\cdots,2\}$ with $A_{j_1},\cdots,A_{j_k}\neq 0$, 
where 
\begin{subequations}
\begin{align}
&E_2=2c_2\widetilde{f}^z+c_4\widetilde{\Theta} A_{-2}^*/A_2,
\label{eq:SM_2}\\
&E_1=c_2\widetilde{f}^z-c_4\widetilde{\Theta} A_{-1}^*/A_1,
\label{eq:SM_1}\\
&E_0=c_4\widetilde{\Theta} A_0^*/A_0,
\label{eq:SM_0}\\
&E_{-1}=-c_2\widetilde{f}^z-c_4\widetilde{\Theta} A_1^*/A_{-1},
\label{eq:SM_-1}\\
&E_{-2}=-2c_2\widetilde{f}^z+c_4\widetilde{\Theta} A_2^*/A_{-2}.
\label{eq:SM_-2}
\end{align}
\end{subequations}
Here, 
$\widetilde{f}^i=
\sum_{\alpha,\beta} A_\alpha^* \mathsf{f}_{\alpha\beta}^i A_\beta$ 
and $\widetilde{\Theta}=2A_2A_{-2}-2A_1A_{-1}+A_0^2$ for simplicity. 

We summarize the result of 
the examination of the consistency conditions. 
It is sufficient to specify amplitudes by their representatives 
by virtue of the SU(2) symmetry. 
\begin{itemize}
\item
The ferromagnetic states, 
$|\widetilde{\bm{f}}|>0$ and $|\widetilde{\Theta}|\ge 0$. 
\begin{alignat}{2}
&\bm{A}=(p_2,0,0,0,p_{-2}),&\qquad& 4c_2=c_4. \\
&\bm{A}=(0,p_1,0,p_{-1},0),&\qquad& c_2=c_4. \\
&\bm{A}=(p_2,p_1,p_0,p_{-1},p_{-2}),&\qquad& c_2=0,\ \widetilde{\Theta}=0.
\label{eq:SM_ferro-3}
\end{alignat}
\item
The polar states, 
$\widetilde{\bm{f}}=0$ and $|\widetilde{\Theta}|>0$.
\begin{alignat}{3}
&\bm{A}=(p_2,0,p_0,0,p_{-2}),&\qquad& |p_2|=|p_{-2}|,\ c_4=0.
\label{eq:polar-1}\\
&\bm{A}=(p,q,r,-q^*,p^*),&\qquad& p,q\in\mathbb{C},\ r\in\mathbb{R}.
\label{eq:polar-2}
\end{alignat}
\item
The cyclic state, 
$\widetilde{\bm{f}}=0$ and $|\widetilde{\Theta}|=0$.
\begin{align}
\bm{A}=(p,0,0,\sqrt{2}q,0),\qquad p,q\in\mathbb{C},\ |p|=|q|.
\end{align}
\end{itemize}
Here we can take $p_i\in\mathbb{C}$. 
These three spin states for $F=2$ spinor BEC are specified in Ref. \citen{CYHo}. 
The cyclic state, which is absent in the $F=1$ case and
is available for spin $F\ge 2$ boson systems, exhibits unusual features
such as phase-locking phenomena and kink excitations \cite{PM}
owing to a unique nature that the condensate
energy depends on the relative value among the phase factors of $\Phi_j$. 
It is remarkable that any one-solitons in the three states are obtained. 
The coefficient of the nonlinear term in (\ref{eq:1NLSE}) turns out to be 
\begin{alignat}{3}
&C=c_0, &\qquad &\mbox{for }\widetilde{\Theta}=0,\\
&C=c_0+c_4, &\qquad &\mbox{otherwise}.
\end{alignat} 

When the effective coupling is the attractive one, 
$C<0$, 
we have the bright one-soliton
\begin{align}
\phi(x,t)=
\frac{\sqrt{2}k_i}{\sqrt{|C|}}\e^{\im\chi_i}\mathrm{sech}\chi_r .
\end{align}
The position function $\chi_r$ and the phase function $\chi_i$
of the soliton are given by 
\begin{align}
&\chi_r=2k_rk_it-k_ix+\delta_r,
\label{eq:chi_r}\\
&\chi_i=-(k_r^2-k_i^2)t+k_rx+\delta_i, 
\label{eq:chi_i}
\end{align}
respectively, where $\chi\equiv\chi_r+\im\chi_i=kx-k^2t+\delta$ 
with $k=k_r+\im k_i$. 
One can also see the plane-wave solution 
$\phi(x,t)=\exp[\im(Kx-\Omega t)]$, where 
$K$ and $\Omega$ are real with $\Omega=K^2+C$. 
Under $C<0$, this plane-wave is unstable against the modulation and 
is decomposed into bright solitons during time-evolution. 
In the case of the repulsive coupling $C>0$, a dark-soliton is formed 
under the nonvanishing boundary conditions $|\phi|\to \mathrm{const.}$ 
as $x\to\pm\infty$. 

We remark on the reduction to $F=1$. 
If $c_4=0$ and $\F=(0,\f_1',\f_0'/\sqrt{3},\f_{-1}',0)$ 
such that $\f_1^{\prime *}\f_0'+\f_0^{\prime *}\f_{-1}'=0$, 
the GP equation for $\f_j'$ is equivalent to the $F=1$ GP equation. 
The polar single-mode soliton with (\ref{eq:polar-2}) and $p=0$ is 
reduced to the $F=1$ polar one-soliton in Refs. \citen{IedaMW1,IedaMW2,WadatiT}.

\section{Hirota's Direct Method}
\label{sec:Hirota}

In this section, we introduce Hirota's direct method \cite{Hirota}. 
Hirota's direct method is powerful for getting solitons 
in both integrable and nonintegrable one-component 
partial differential equation (PDE) systems. 
We generalize this method for the $F=2$ GP equation. 

By putting $\f_j=G_j/H$ for $j=-2,\cdots,2$, 
the GP equation is transformed into the form, 
\begin{subequations}
\label{eq:bilinear}
\begin{align}
&(\im D_t+D_x^2)G_j\cdot H
-\left(\frac{c_2}{2}\frac{\delta\bm{f}^2}{\delta\f_j^*}
+\frac{c_4}{2}\frac{\delta|\Theta|^2}{\delta\f_j^*}\right)H^2=0,
\label{eq:bilinear-1}\\
&D_x^2 H\cdot H+c_0\sum_{\alpha=-2,\cdots,2}|G_\alpha|^2=0.
\label{eq:bilinear-2}
\end{align}
\end{subequations}
Here the Hirota derivative is defined as 
\begin{align}
D_t^mD_x^n a\cdot b=
\left(\frac{\6}{\6 t_1}-\frac{\6}{\6 t_2}\right)^m
\left(\frac{\6}{\6 x_1}-\frac{\6}{\6 x_2}\right)^n
a(x_1,t_1)b(x_2,t_2)\bigg|_{{{x_1=x_2=x}\atop{t_1=t_2=t}}}.
\end{align}
In the Manakov and the one-component systems, 
the $c_2$- and $c_4$-terms 
in eqs. (\ref{eq:bilinear-1}) are absent and 
the transformation is called the bilinear transformation. 

For attractive spin-independent interaction, 
we can set $c_0=-2$ by scaling the order parameters. 
Equations (\ref{eq:bilinear}) give solitons 
with the vanishing boundary conditions $|\Phi|\to 0$ as $x\to\pm\infty$, 
that is, bright solitons.
For the nonvanishing boundary conditions, extra terms are needed. 

Within Hirota's method, 
a one-soliton solution may be obtained by a finite-order perturbation, 
\begin{subequations}
\label{eq:Hirota-HG}
\begin{align}
&H=1+\varepsilon^2 h_2+\varepsilon^4 h_4,
\label{eq:Hirota-H}\\
&G_j=\varepsilon g_{1,j}+\varepsilon^3 g_{3,j}.
\label{eq:Hirota-G}
\end{align}
\end{subequations}
Substituting (\ref{eq:Hirota-HG}) into (\ref{eq:bilinear}), 
we solve them order by order. 
At the first order, we have 
\begin{align}
g_{1,j}=\Pi_j\e^{\im(kx-k^2t)}
\label{eq:g1}
\end{align}
with 
$k=k_r+\im k_i\in \mathbb{C}$ and 
$\Pi_j$ ($j=-2,\cdots,2$) being free parameters for one-solitons. 
A success of terminating the perturbation expansion at a finite order 
leads to soliton solutions. 
We cannot expect soliton solutions for generic values of parameters. 
Our strategy is that by utilizing the freedom of 7 parameters, 
\emph{i.e.} 
5 for one-soliton $\Pi_j$ and 2 for interaction couplings $c_2$ and $c_4$, 
we look for both one-soliton solutions and 
the valid parameters in order to have those solutions. 

It is known that for one-component PDEs, 
Hirota's direct method is applicable for upto 
two-solitons even in nonintegrable systems \cite{Hietarinta}. 
In fact, the expansion (\ref{eq:Hirota-HG}) is usually 
taken for a two-soliton. 
In our case, as we will show later, we have a twin-peak one-soliton 
for the $F=2$ GP equation. 
In a sense, a twin-peak one-soliton may be considered as 
a degenerate two-soliton. 
This kind of observation is also seen in 
Refs. \citen{UchiIW-dark,IedaUchiW-ISM}. 
For a general two-soliton, 
we 
take (\ref{eq:g1}) as a combination of two plane waves. 
In principle, those calculations which include 
more higher orders and multi-solitons are possible 
in integrable systems, 
but we do not reach the argument of integrability and 
postpone them to future works.

\section{Results for One-Solitons}
\label{sec:soliton}

We present one-soliton solutions through Hirota's direct method. 
A one-soliton is determined by the following parameters; 
$k_r$: (half of) the phase velocity of the envelope soliton, 
$k_i$: the amplitude of the soliton, 
$\bm{\Pi}=(\Pi_2,\Pi_1,\Pi_0,\Pi_{-1},\Pi_{-2})$: 
distribution among spin components. 
The position function and the phase function of the soliton are 
the same as the one-component one's, 
(\ref{eq:chi_r}) and (\ref{eq:chi_i}), respectively. 
We normalize $\bm{\Pi}$ by $\sum_j|\Pi_j|^2=1$ 
since its factor can be absorbed in the shift of $\chi$.

\subsection{Single-mode (Single-peak) soliton}

Single-mode one-solitons are reproduced 
with specific values of coupling constants. 


\noindent
\emph{a}) A single-mode soliton in the ferromagnetic state, 
with $|\bm{f}|>0$ and $\Theta=0$. 
\begin{align}
\begin{cases}
&\bm{\Pi}=(p_2,p_1,p_0,p_{-1},p_{-2}),\qquad 
p_i\in\mathbb{C},\ 2p_2p_{-2}-2p_1p_{-1}+p_0^2=0. \\
&c_2=0,\ c_4=\mbox{arbitrary}. 
\end{cases}
\end{align}

\noindent
\emph{b}) A single-mode soliton in the polar state, 
with $\bm{f}=0$ and $|\Theta|>0$. 
\begin{align}
\begin{cases}
&\bm{\Pi}=(p_2,0,p_0,0,p_{-2}),\qquad p_i\in\mathbb{C},\ |p_2|=|p_{-2}|.\\
&\bm{\Pi}=(p,q,r,-q^*,p^*),\qquad p,q\in\mathbb{C},\ r\in\mathbb{R}.\\
&c_2=\mbox{arbitrary},\ c_4=0.
\end{cases}
\end{align}

\noindent
\emph{c}) A single-mode soliton in the cyclic state, 
with $\bm{f}=0$ and $\Theta=0$. 
\begin{align}
\begin{cases}
&\bm{\Pi}=(p,0,0,\sqrt{2}q,0),\qquad p,q\in\mathbb{C},\ |p|=|q|.\\
&c_2=\mbox{arbitrary},\ c_4=\mbox{arbitrary}. 
\end{cases}
\end{align}
For the cases \emph{a})$\sim$\emph{c}), the one-soliton solution is 
\begin{align}
\F=k_i\e^{\im\chi_i}\mathrm{sech}\chi_r\bm{\Pi}.
\label{eq:sech-1}
\end{align}
It should be remarked that the cyclic soliton in \emph{c}) exists 
for all interaction couplings.

\noindent
\emph{d}) A single-mode soliton in the polar state, 
with $\bm{f}=0$, $|\Theta|>0$ and 
a doubled particle number compared to the cases \emph{a})$\sim$\emph{c}) 
for the same $k_r$ and $k_i$. 
\begin{align}
\begin{cases}
&\bm{\Pi}=(p,q,r,-q^*,p^*),\qquad 
p,q\in\mathbb{C},\ r\in\mathbb{R}, \\
&\bm{\Pi}=(p_2,0,p_0,0,p_{-2}),\qquad 
p_i\in\mathbb{C},\ 
|p_2|=|p_{-2}|,\ p_0^2/p_2p_{-2}\in\mathbb{R}.\\
&c_2=\mbox{arbitrary},\ c_4=1. 
\end{cases}
\end{align}

\noindent
\emph{e}) A single-mode soliton in the ferromagnetic state, 
with $|\bm{f}|>0$, $|\Theta|\ge 0$ and 
a doubled particle number compared to the cases \emph{a})$\sim$\emph{c}) 
for the same $k_r$ and $k_i$. 

\begin{align}
&
\begin{cases}
&\bm{\Pi}=(p_2,0,0,0,p_{-2}),\qquad p_i\in\mathbb{C}.\\
&c_2=1/4,\ c_4=1. 
\end{cases}
\\
&
\begin{cases}
&\bm{\Pi}=(p,0,0,0,0),\qquad p\in\mathbb{C}.\\
&c_2=1/4,\ c_4=\mbox{arbitrary}. 
\end{cases}
\\
&
\begin{cases}
&\bm{\Pi}=(0,p_1,0,p_{-1},0),\qquad p_i\in\mathbb{C}.\\
&c_2=1,\ c_4=1. 
\end{cases}
\\
&
\begin{cases}
&\bm{\Pi}=(0,p,0,0,0),\qquad p\in\mathbb{C}.\\
&c_2=1,\ c_4=\mbox{arbitrary}. 
\end{cases}
\end{align}
For the cases \emph{d}) and \emph{e}), the one-soliton solution is 
\begin{align}
\F=\sqrt{2}k_i\e^{\im\chi_i}\mathrm{sech}\chi_r\bm{\Pi}.
\end{align}
Compared to (\ref{eq:sech-1}) in the previous cases, 
the amplitude is larger by a factor $\sqrt{2}$. 
From the single-mode analysis in Sec. \ref{sec:SM}, 
it is shown that the effective coupling $C$ for the one-component equation 
has the value $c_0+c_4=-1$, not $c_0=-2$. 
Accordingly, the amplitude gets the factor $\sqrt{2}$, 
and hence the particle number becomes doubled. 

The single-mode solitons in this section are included in 
the result in Sec. \ref{sec:SM}. 

One can see that for single-mode one-solitons, 
zero local spin is allowed for arbitrary $c_2$, and 
zero singlet-pair amplitude is allowed for arbitrary $c_4$, 
since the corresponding interaction energy in (\ref{eq:GPenergy}) is 
ineffective, respectively. 
Moreover, the solitons in the cases \emph{a})$\sim$\emph{c}) turn to have 
vanishing interaction terms with $c_2$ and $c_4$ 
in the energy (\ref{eq:GPenergy}), and 
they are regarded as the solitons in the Manakov system.

\subsection{Twin-peak soliton}

We further investigate one-solitons which cannot be expressed 
within a single-mode form. 
They have a wave-form with twin peaks. 
The distance of the twin peaks is freely adjusted by changing 
the parameters of the one-soliton. 
One may be tempted to say that the twin-peak soliton is the superposition of 
two identical single-peak solitons with shift of their positions, 
but it is not true because 
physical densities of the twin-peak soliton are not always just the 
sum of those of two single-peak solitons. 
Such a twin-peak one-soliton was already discovered in the $F=1$ 
spinor BEC \cite{IedaMW1,IedaMW2}, 
but is allowed only for the polar state, 
\emph{i.e.} with zero total spin. 
In our result for $F=2$, we find that twin-peak one-soliton occurs 
both in the polar state and in the ferromagnetic state.

\noindent
\emph{f}) A twin-peak soliton in the polar state, 
with $\bm{F}=0$ (but locally $\bm{f}\neq 0$) and $|\Theta|>0$. 
\begin{align}
\begin{cases}
&\bm{\Pi}=(p_2,p_1,p_0,p_{-1},p_{-2}),\qquad p_i\in\mathbb{C}. \\
&c_2=0,\ c_4=1. 
\end{cases}
\end{align}
The wavefunctions have the form 
\begin{align}
\f_j=\frac{\sqrt{2}k_i}{\sqrt{|\T|}}
\frac{(-1)^j\sigma p_{-j}^*\e^{\chi_r}
	+p_j\e^{-\chi_r}}
	{\cosh 2\chi_r+\cosh\omega}
\e^{\im\chi_i}, 
\end{align}
for $j=-2,\cdots,2$, 
where $\T=2p_2p_{-2}-2p_1p_{-1}+p_0^2$, $\N=\sum_j |p_j|^2$, 
$\sigma=\T/|\T|$ and $\cosh\omega=\N/|\T|$. 
Physical densities are calculated as follows; 
\begin{align}
&n=k_i^2\left[
\mathrm{sech}^2\left(\chi_r-\frac{\omega}{2}\right)
+\mathrm{sech}^2\left(\chi_r+\frac{\omega}{2}\right)
\right],
\\
&f^i=-4k_i^2\frac{\mathfrak{f}^i}{|\T|}
\frac{\sinh 2\chi_r}{(\cosh 2\chi_r+\cosh\omega)^2},
\\
&|\Theta|=\frac{4k_i^2}{\cosh 2\chi_r+\cosh\omega},
\end{align}
where 
$\mathfrak{f}^i=\sum_{\alpha,\beta} 
\Pi_\alpha^* \mathsf{f}^i_{\alpha\beta}\Pi_\beta$. 
The total amounts of the quantities are obtained by integrating 
these densities as 
\begin{align}
&N=4k_i,\\
&\bm{F}=(0,0,0),\\
&S=4k_i\omega\ \mathrm{cosech}\omega.
\end{align}
\begin{figure}[t]
\begin{center}
\begin{minipage}{7cm}
\includegraphics[bb=0 0 234 185,scale=0.8]{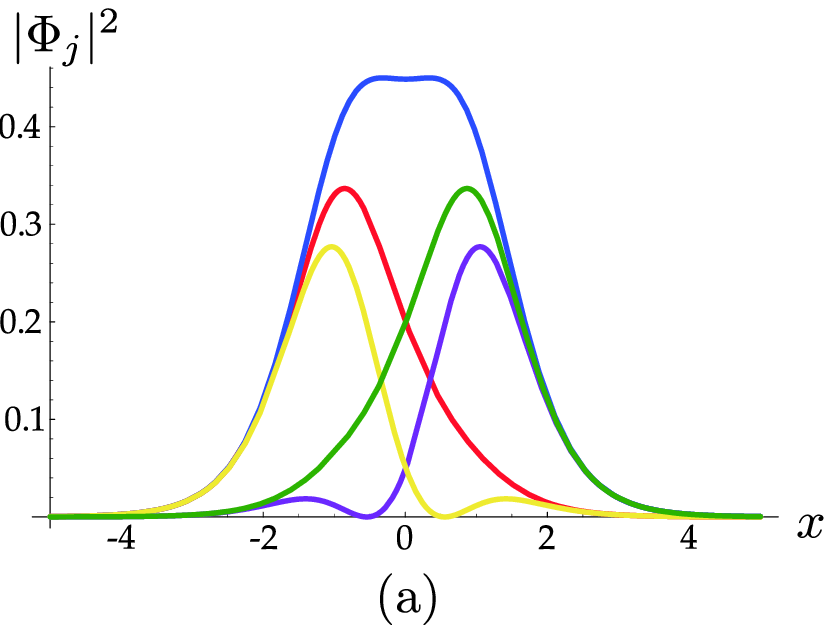}
\end{minipage}
\begin{minipage}{7cm}
\includegraphics[bb=0 0 234 184,scale=0.8]{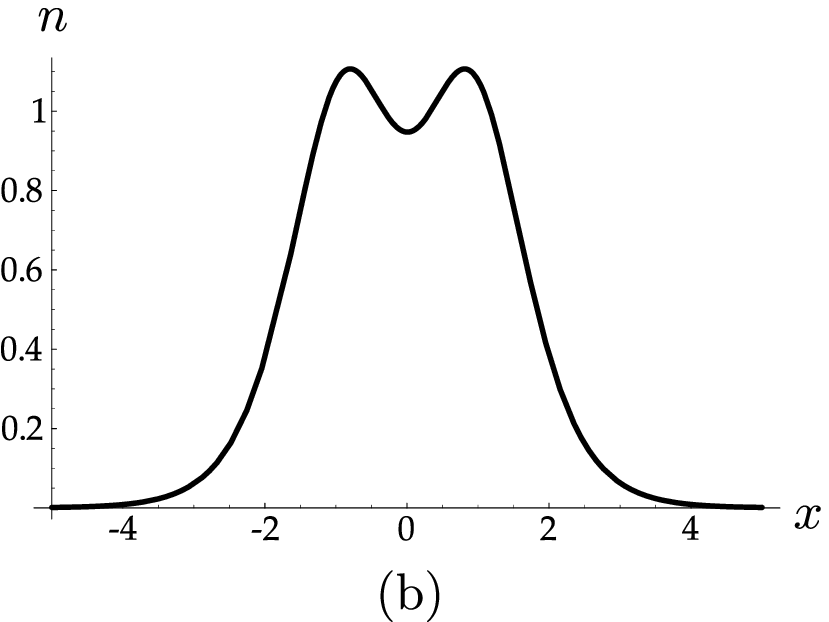}
\end{minipage}
\begin{minipage}{7cm}
\includegraphics[bb=0 0 230 181,scale=0.8]{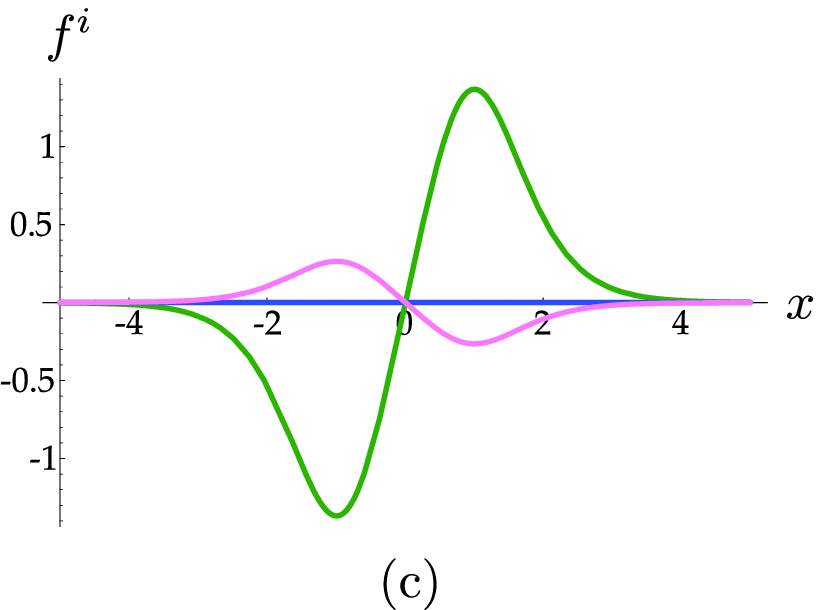}
\end{minipage}
\begin{minipage}{7cm}
\includegraphics[bb=0 0 234 193,scale=0.8]{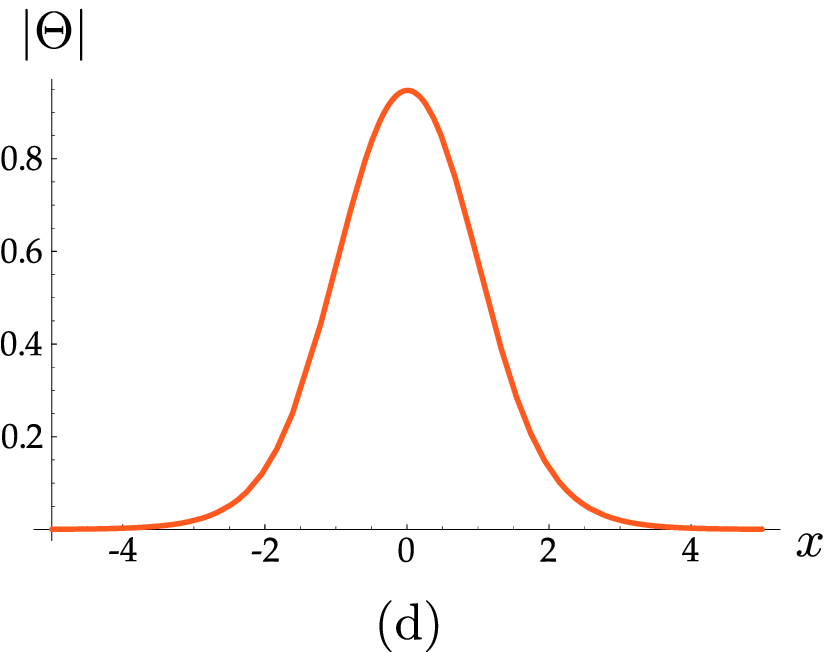}
\end{minipage}
\caption{Density plots for \emph{f}) (twin-peak polar soliton) 
with $\bm{\Pi}=(1,3,3,1,3)$, $k_r=k_i=1$. 
(a) Densities for each component 
($|\f_2|^2$: red, $|\f_1|^2$: purple, $|\f_0|^2$: blue, 
$|\f_{-1}|^2$: yellow, $|\f_{-2}|^2$: green). (b) The number density. 
(c) The spin densities ($f^x$: green, $f^y$: blue, $f^z$: pink). 
(d) The absolute value of the singlet-pair amplitude. 
\label{fig:twin_polar}
}
\end{center}
\end{figure}

Figure \ref{fig:twin_polar} shows an example of the twin-peak polar soliton. 
We observe that the spins contained in the two peaks 
have the same amount with the opposite sign, and form a polarization. 
Therefore, in total the soliton has zero spin. 
The reduction to a single-peak soliton is achieved by 
sending $\T\to 0$. 
In this limit, the two peaks get infinitely far apart and 
eventually, the remained single-peak soliton 
contains nonzero total spin and no singlet-pair amplitude, 
and coincides with the ferromagnetic one (\ref{eq:SM_ferro-3}) in \emph{a}). 
We also observe that the singlet-pair amplitude is localized around 
the center of the twin peaks. 
This indicates that the twin-peak soliton is not just the superposition 
of two identical single-peak solitons. 
The same argument also holds for the next case \emph{g}). 

For the special case with zero local spin, 
where we make $\mathfrak{f}^i=0$ for $i=x,y,z$, 
the following set of parameters is allowed: 
\begin{align}
\begin{cases}
&\bm{\Pi}=(p_2,0,p_0,0,p_{-2}),\qquad p_i\in\mathbb{C},\ |p_2|=|p_{-2}|. \\
&c_2=\mbox{arbitrary},\ c_4=1. 
\end{cases}
\end{align}
The wavefunctions are 
\begin{subequations}
\begin{align}
&\f_{\pm 2}=
\frac{\sqrt{2}k_ip_{\pm 2}}{(2p_2p_{-2}+p_0^2)^2}
\frac{\e^{\chi_r+\im\xi}+\e^{-\chi_r}}
	{\cosh 2\chi_r+ \cosh\omega}\e^{\im\chi_i},\\
&\f_{\pm 1}=0,\\
&\f_0=
\frac{\sqrt{2}k_ip_0}{(2p_2p_{-2}+p_0^2)^2}
\frac{\e^{\chi_r+\im\xi'}+\e^{-\chi_r}}
	{\cosh 2\chi_r+ \cosh\omega}\e^{\im\chi_i},
\end{align}
\end{subequations}
where $\cosh\omega=\frac{2|p_2p_{-2}|+|p_0|^2}{|2p_2p_{-2}+p_0^2|}$, 
$\xi=\arg \left(1+\frac{p_0^2}{2p_2p_{-2}}\right)$ and 
$\xi'=\arg \left(1+\frac{2p_2p_{-2}}{p_0^2}\right)$.
In particular, $\xi=\xi'(=0)\ (\cosh\omega =1)$ gives 
the single-mode soliton of \emph{d}).

\noindent
\emph{g}) A twin-peak soliton in the ferromagnetic state, 
with $|\bm{f}|>0$ and $|\Theta|>0$. 
\begin{align}
&
\begin{cases}
&\bm{\Pi}=(p_+,0,0,0,p_-),\qquad p_i\in\mathbb{C}.\\
&c_2=1/4,\ c_4=0. 
\end{cases}
\label{eq:twin-2-1}\\
&
\begin{cases}
&\bm{\Pi}=(0,p_+,0,p_-,0),\qquad p_i\in\mathbb{C}. \\
&c_2=1,\ c_4=0. 
\label{eq:twin-2-2}
\end{cases}
\end{align}
We use $s=2$ for the case (\ref{eq:twin-2-1}) and 
$s=1$ for the case (\ref{eq:twin-2-2}). 
Then, the wavefunctions are 
\begin{align}
\f_{\pm s}=\frac{\sqrt{2}k_ip_\pm}{\sqrt{||p_+|^2-|p_-|^2|}}
\frac{\pm\sigma\e^{\chi_r}+\e^{-\chi_r}}{\cosh 2\chi_r+\cosh\omega}
\e^{\im\chi_i},
\label{eq:phi_twin-ferro}
\end{align}
and the others are constantly zero, 
where $\sigma=\mathrm{sign}\left(|p_+|^2-|p_-|^2\right)$ and 
$\cosh\omega=\frac{|p_+|^2+|p_-|^2}{||p_+|^2-|p_-|^2|}$. 
The densities are given by 
\begin{align}
&n=k_i^2\left[
\mathrm{sech}^2\left(\chi_r-\frac{\omega}{2}\right)
+\mathrm{sech}^2\left(\chi_r+\frac{\omega}{2}\right)
\right],
\\
&f^x=f^y=0,\\
&f^z=\frac{4k_i^2s\sigma}{\cosh 2\chi_r+\cosh\omega},
\\
&|\Theta|=\frac{4k_i^2|\sinh\omega \sinh 2\chi_r|}
{(\cosh 2\chi_r+\cosh\omega)^2}.
\end{align}
The total amounts of the quantities are 
\begin{align}
&N=4k_i,\\
&\bm{F}=(0,0,4k_is\sigma\omega\ \mathrm{cosech}\omega),\\
&S=4k_i\left|\frac{p_-}{p_+}\right|^\sigma.
\end{align}

\begin{figure}[t]
\begin{center}
\begin{minipage}{7cm}
\includegraphics[bb=0 0 234 193,scale=0.8]{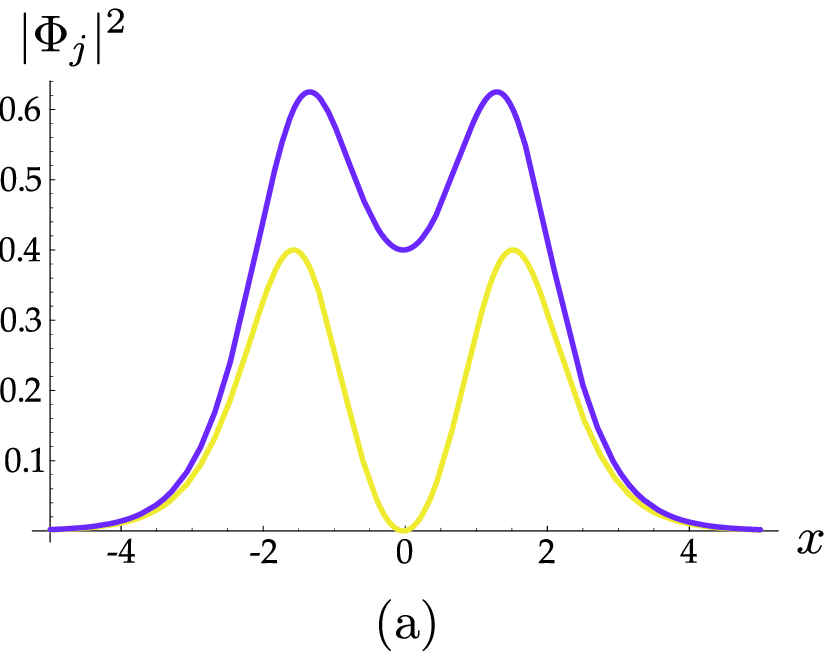}
\end{minipage}
\begin{minipage}{7cm}
\includegraphics[bb=0 0 234 193,scale=0.8]{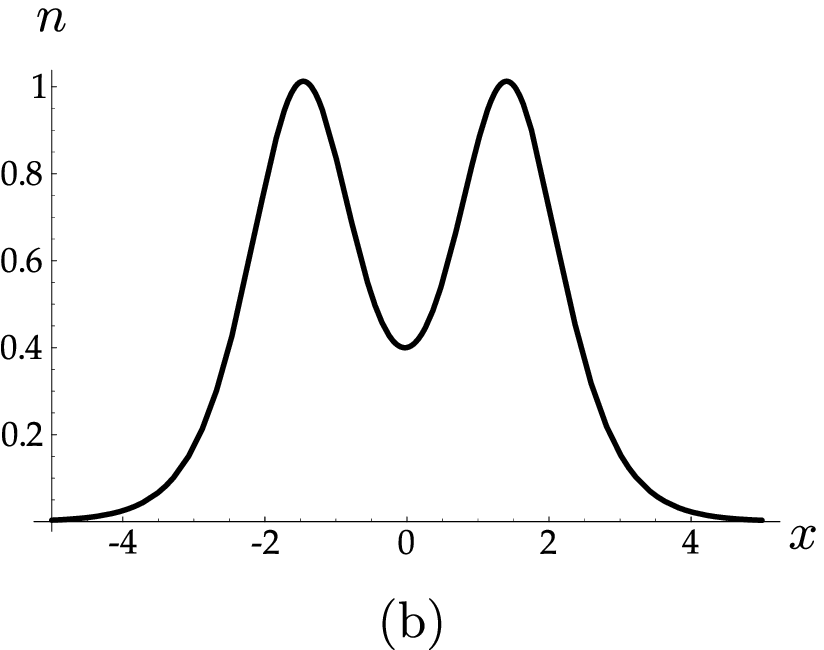}
\end{minipage}
\begin{minipage}{7cm}
\includegraphics[bb=0 0 234 197,scale=0.8]{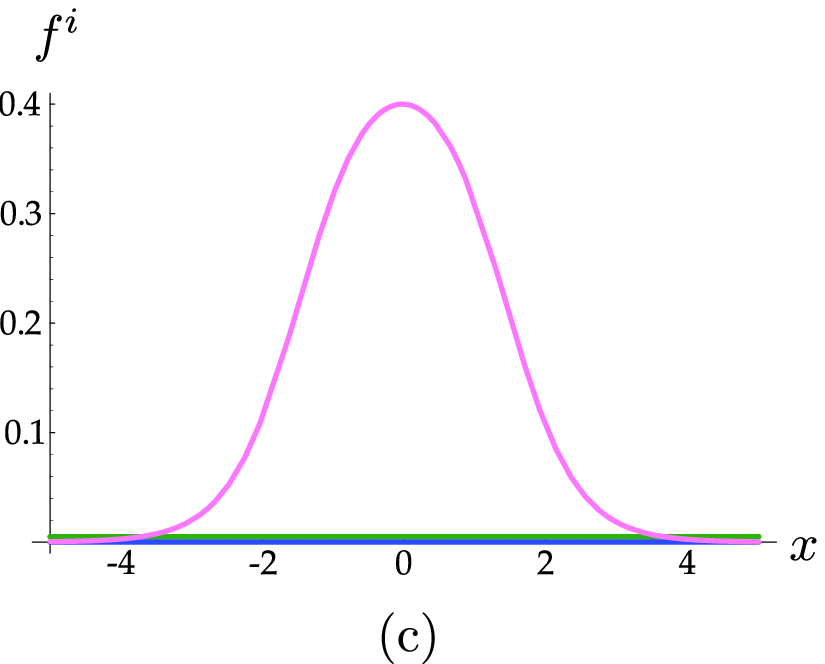}
\end{minipage}
\begin{minipage}{7cm}
\includegraphics[bb=0 0 234 196,scale=0.8]{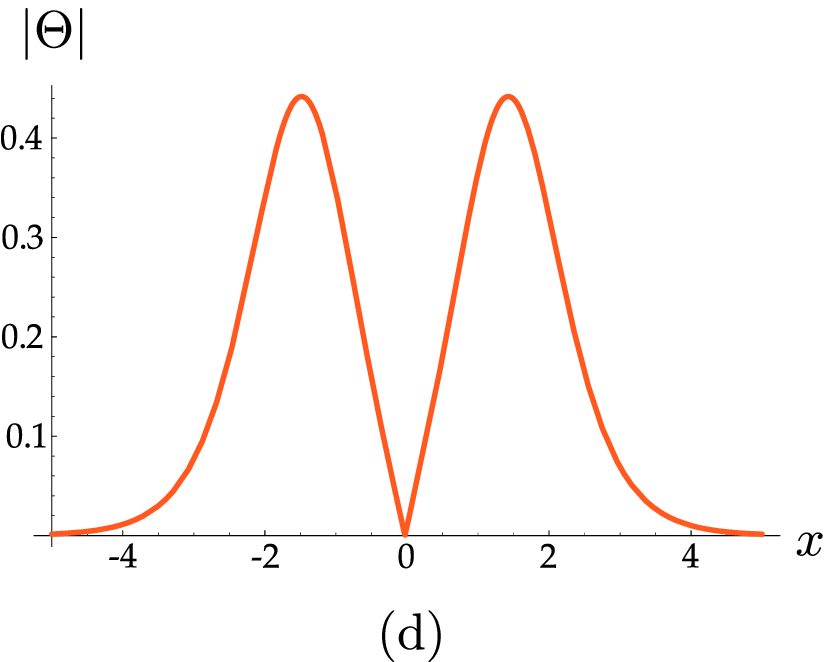}
\end{minipage}
\caption{Density plots for \emph{g}) (twin-peak ferromagnetic soliton) 
with $\bm{\Pi}=(0,\sqrt{5}/2,0,1,0)$, $k_r=k_i=1$. 
(a) Densities for each component ($|\f_1|^2$: purple, $|\f_{-1}|^2$: yellow). 
(b) The number density. 
(c) The spin densities ($f^x$: green, $f^y$: blue, $f^z$: pink). 
(d) The absolute value of the singlet-pair amplitude. 
\label{fig:twin_ferro}
}
\end{center}
\end{figure}

Figure \ref{fig:twin_ferro} illustrates 
an example of the twin-peak ferromagnetic soliton. 
One can see that 
the spin densities are localized around the center of the twin peaks. 
As $|p_-|/|p_+|\to 1$, the two peaks of the soliton get infinitely far apart and 
a single-peak soliton in the polar state 
with (\ref{eq:polar-1}) in \emph{b}) is left. 
It is interesting that the reduction of the twin-peak soliton 
to the single-peak one 
changes the state from ferromagnetic to polar in \emph{g})
and vice versa in \emph{f}). 

Note that (\ref{eq:phi_twin-ferro}) is the solution of 
the two-component coupled NLSE \cite{cNLSE1}, 
\begin{subequations}
\begin{align}
&\mathcal{L}\f_s=
-\left(\alpha|\f_s|^2+\beta|\f_{-s}|^2\right)\f_s,\\
&\mathcal{L}\f_{-s}=
-\left(\beta|\f_s|^2+\alpha|\f_{-s}|^2\right)\f_{-s},
\end{align}
\end{subequations}
with $\alpha=1$ and $\beta=3$. 
It was established that 
two-component coupled NLSE is integrable only for $\alpha=\beta$, 
corresponding to the original Manakov equation \cite{ZakSch}.

\section{Discussion and Conclusion}
\label{sec:concl}

We have studied one-soliton solutions for 
the Gross--Pitaevskii (GP) equation of 
the $F=2$ spinor Bose--Einstein 
condensates (BEC) by means of two methods, the single-mode analysis 
and the multi-component generalization of Hirota's direct method. 
The latter method has been successfully applied to show twin-peak solitons 
both in the ferromagnetic and the polar states, 
which cannot be accessed by the single-mode analysis. 

Hirota's method is not restricted to the present analysis. 
One can also find soliton solutions for higher-spinor BECs or 
other types of multi-component systems by generalizing this method 
as presented in this work. 
For instance, applying to the $F=1$ spinor BEC, 
we reproduce the bright solitons in 
Refs. \citen{IedaMW1,IedaMW2,WadatiT}, 
at the order $\varepsilon^2$ for the single-mode solitons and 
at the order $\varepsilon^4$ for the twin-peak polar soliton. 
Mathematically, incorporating such as $c_2$- and $c_4$-terms 
of eqs. (\ref{eq:bilinear-1}) in Hirota's method suggests a new direction 
for further extension of the framework. 

One of our next interests is the integrability of the GP equation. 
In the integrable systems with multi-components, 
multi-solitons should be constructed from any combinations of 
one-solitons. 
Their collisions are factorized into 
successive two-soliton collisions, but 
in contrast to one-component systems, it is not always the case that 
each soliton keeps its shape after collisions, 
deforming its parameters for internal degrees of freedom. 
The results in this paper pick up specific interactions, 
and we hope that they especially include integrable points. 
Whether those systems are integrable or equivalent to 
already known systems is an interesting future problem. 
The Painlev\'e analysis may give a clue for the problem. 

From the physical point of view, 
the discovery of all one-solitons 
in the ferromagnetic, polar and cyclic states 
is of much importance. 
As higher-spinor BECs should exhibit richer physics, 
we expect that higher-spinor solitons will make wider 
possibilities in various applications.

%
%

\end{document}